\documentclass[12pt]{article}
\usepackage{amsmath,amssymb,amsthm,latexsym}
\usepackage{stmaryrd,wasysym,upgreek,mathrsfs,dsfont}
\usepackage{graphicx,color}

\usepackage{tensor}
\usepackage{verbatim}
\usepackage{epigraph}
\usepackage{url}
\usepackage{physics}
\usepackage{todonotes}
\usepackage{wasysym} 
\usepackage{marvosym} 
\usepackage{textcomp} 
\usepackage[colorlinks=true, allcolors=blue]{hyperref}
\usepackage{footmisc}

\newcommand{\beq}{\begin{equation}}
\newcommand{\eeq}{\end{equation}}
\newcommand{\be}{\begin{equation}}
\newcommand{\bee}{\begin{equation}}
\newcommand{\ee}{\end{equation}}
\newcommand{\bea}{\begin{eqnarray}}
\newcommand{\eea}{\end{eqnarray}}

\newtheorem{definition}{Definition}[section]
\newtheorem{prop}{Proposition}[section]
\newtheorem{lemma}[prop]{Lemma}

\newtheorem{theorem}[prop]{Theorem}
\newtheorem{cor}[prop]{Corollary}

\newcommand{\bal}{\begin{align}}
\newcommand{\eal}{\end{align}}

\newcommand{\cC}{{\cal{C}}}

\newcommand{\cE}{{\cal{E}}}

\newcommand{\cL}{{\cal{L}}}

\newcommand{\cT}{{\cal{T}}}

\newcommand{\cF}{{\cal{F}}}
\newcommand{\cS}{{\cal{S}}}
\newcommand{\cU}{{\cal{U}}}

\newcommand{\R}{\mathbb{R}}

\newcommand{\N}{\mathbb{N}} 
\newcommand{\mP}{{\mathbb{P}}}
\newcommand{\mE}{{\mathbb{E}}}

\newcommand{\prf}{{\noindent {\rm \bf Proof}\, }}

\newcommand{\bbbone}{{\mathds{1}}}

\newcommand{\rmq}[1]{{\color{red}#1}}

\begin{document}

\title{Perturbative Quantum Field Theory\\ on  Random Trees}

\author{Nicolas Delporte 
and Vincent Rivasseau\\
        Laboratoire de physique th\'eorique\\
        CNRS UMR6827 and  Universit\'e Paris-Sud,\\ Universit\'e Paris-Saclay, 91405 Orsay, France\\ \small{{\tt nicolas.delporte@th.u-psud.fr} ; {\tt vincent.rivasseau@th.u-psud.fr}}}


\maketitle

\begin{abstract} In this paper we start a systematic study of quantum field theory on random trees.
Using precise probability estimates on their Galton-Watson branches and a multiscale analysis,
we establish the general power counting of averaged Feynman amplitudes and check that they behave indeed
as living on an effective space of dimension 4/3, the spectral dimension of random trees. 
In the ``just renormalizable" case 
we prove convergence of the averaged amplitude of any completely convergent graph, and
establish the basic localization and subtraction estimates required for perturbative renormalization. Possible consequences for an SYK-like model on random trees are briefly discussed.

\end{abstract}
\pagebreak
\tableofcontents

\section{Introduction}

In the Euclidean path integral formulation, 
quantizing gravity translates into randomizing geometry pondered by 
the Einstein-Hilbert action or some generalization thereof. Since a direct continuum
formulation is plagued with many delicate issues, from non-renormalizable ultraviolet divergences to the huge gauge group of diffeomorphism invariance and the impossibility to fully classify geometries in dimension higher than two through complete lists of invariants, the safest road seems to search for generic, sufficiently universal large distance/semi-classical limits of \emph{discretized} random geometries, using both analytic and numerical tools. 
In this approach to quantum gravity, space-time is no longer 
fixed, but sampled from a statistical collection
of large discrete objects such as triangulations or their dual graphs \cite{ambjorn}. 

To understand the physical (potentially observable) consequences of such 
a bold point of view, it is important to study how particles propagate and interact on such statistical collections of random graphs. 

\begin{figure}[ht]
\centerline{\includegraphics[width=14cm]{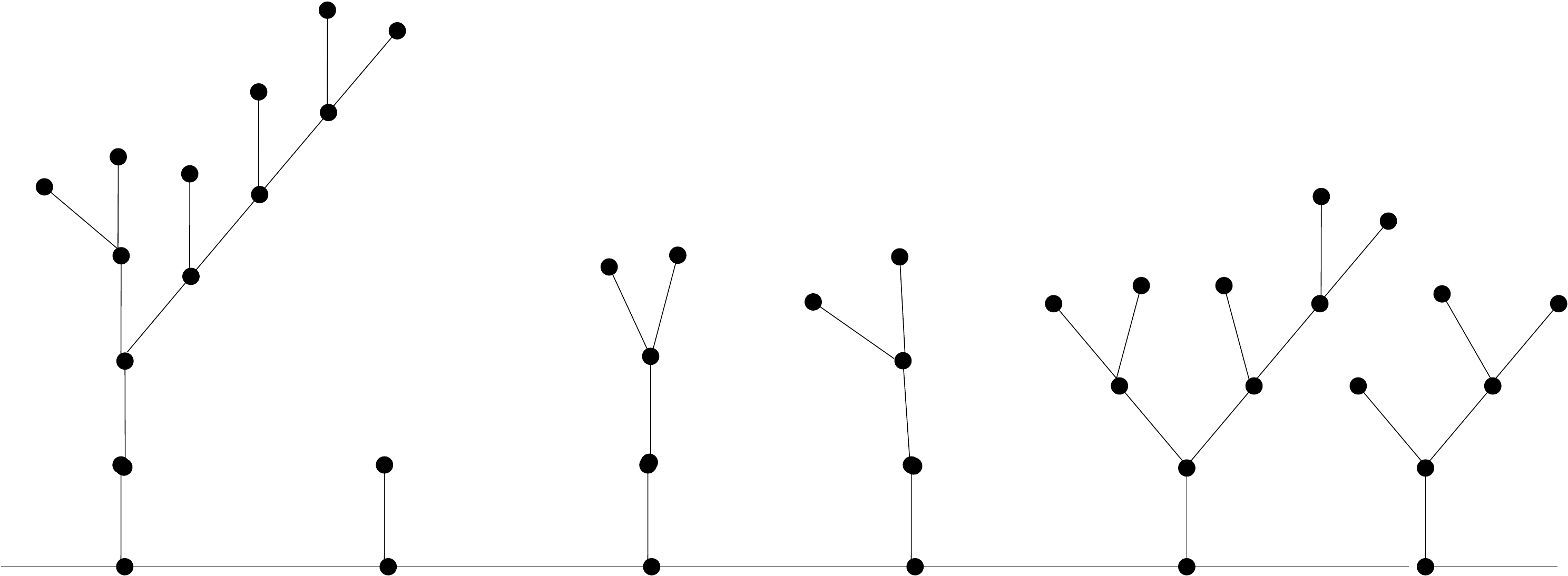}}
\caption{An infinite binary  tree with horizontal spine and Galton-Watson branches}
\label{treeandspine}
\end{figure}

\begin{figure}
\centerline{\includegraphics[width=13cm]{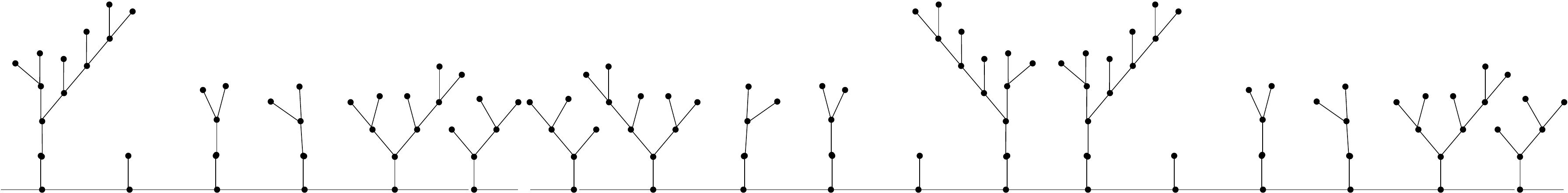}    }
\medskip
\centerline{\Large  $\Downarrow$}
\vskip .5cm
\centerline{\includegraphics[width=13cm]{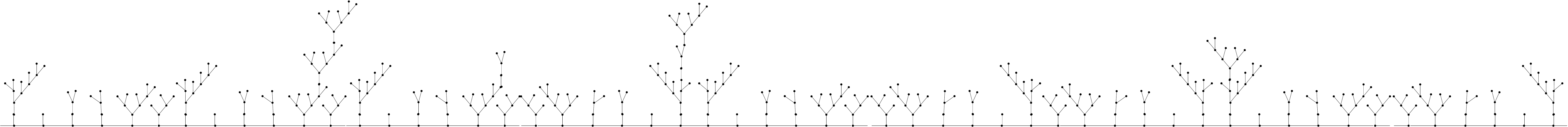}    }
\caption{Zooming towards the continuum random tree}
\label{treezooming}
\end{figure}

Random trees (often known in physics under the name of  branched polymers)
are the first and most natural examples of random graphs. In the large size/continuum limit, they have good universal properties. Mathematicians have been studying them in detail with combinatorial, analytic and probabilistic tools such as the basic map from trees to brownian excursions, Fuss-Catalan numbers 
\cite{FussCatalan}, Galton-Watson processes \cite{harris}  and so on.
There are now fairly detailed rigorous results on their continuum limit \cite{aldous} and universal critical indices such as their Hausdorff dimension ($d_H = 2$), and 
their spectral dimension ($d_S = 4/3$) \cite{DJW,BarlowKumagai,Kumagai}.
An essential characteristic of one of the simplest class of infinite random trees 
considered in the literature \cite{aldous,DJW,BarlowKumagai,Kumagai} is the 
existence of a \emph{single infinite} 
one-dimensional \emph{spine}\footnote{The usual spine obtained by conditioning a critical Galton-Watson tree by non-extinction corresponds 
to a one dimensional \emph{half-space}. However it should be straightforward to symmetrize the spine
to get a full one dimensional space.}, decorated 
by independent random \emph{finite} critical Galton-Watson branches (see Figure \ref{treeandspine}). 
Figure \ref{treezooming} tries to give an intuition 
of zooming towards the large size/continuum limit of random trees.

The more complicated case of two dimensional random geometries 
and quantum gravity \cite{DiFrancesco:1993cyw} is also relatively well understood.
The typical random space here is the now famous 
Brownian sphere \cite{LeGall,Miermont} ($d_H = 4$, $d_S= 2$).
The main result to remember is that this Brownian sphere,
which is the continuum limit of planar $q$-angulations, themselves dual to the dominant
graphs of matrix models with $\Tr M^q$ interaction \cite{HooftBIPZ}, is \emph{the same} \cite{MillerSheffield} as the Liouville quantum field theory formulation of pure two dimensional quantum gravity, where the Liouville field describes the conformal factor of the random metric \cite{Polyakov:1981rd,HJ,DKRV}. Planar graphs (more technically planar combinatorial maps) can be thought of as a natural evolution of random trees through the addition of some random labels as in the Cori-Vauquelin-Schaeffer \cite{CVS} and Bouttier-di-Francesco-Guitter \cite{BFG} bijections, or through equivalent \emph{mating processes} \cite{MatingTrees}.

Quantum field theory on random spaces 
has been developed mostly not on random trees\footnote{See however e.g. \cite{statmechtrees}
for statistical mechanic models on random trees and graphs.} but on the more complicated two dimensional random geometries for many reasons. Physicists are very interested 
in conformal field theories (CFT) since they enjoy universal properties as fixed points of the renormalization group. In flat two dimensional space, there exists a rich family of non-trivial CFTs for which exact analytic results can be obtained. When such CFTs are coupled to Liouville gravity, 
the critical indices of matter are modified in a computable way 
through the celebrated Knizhnik-Polyakov-Zamolodchikov \cite{KPZ} and David-Distler-Kawai \cite{DDK} relations. This led during the last fourty years to a flurry of marvelous results, 
both in theoretical physics and mathematics, that we cannot even roughly sketch here \cite{Duplantier}. The link with mainstream string theory is a powerful motivation  \cite{Polyakov:1981rd}. Also in such studies
the sphere or the $\R^2$ plane still provides a fixed 
background topology. Randomness of space-time is reduced to the  familiar Liouville scalar field which represents the fluctuation of the conformal factor of the metric. Clearly this is conceptually less disturbing than a completely random geometric point of view, 
for which even observables may not be obvious to define, as they have to be attached to
features common to almost all objects of the statistical sum.
Finally and perhaps most importantly, random trees and branched polymers
were considered until rather recently quite trivial and unpromising for quantum gravity. 
This has changed following two discoveries of the last decade.

Firstly the large N expansion of random tensor models \cite{tensors} was found \cite{largeN}.
Melonic graphs dominate at leading order \cite{melonic}. 
The generality and robustness of this result has been confirmed 
more and more with time \cite{robust}.
Since the dual graphs of random tensors of rank $r$
perform a statistical sum over a huge geometric category including all piecewise linear 
quasi-manifolds of dimension $r$, the melonic graphs form a natural entrance door 
to higher dimensional quantum gravity, a point of view advocated in
\cite{tensortrack}. However melons are a strict subset of planar graphs.
Equipped with the graph distance, they have as scaling limit the
branched polymer phase \cite{Gurau:2013cbh}. This seems a puzzling 
step backward compared to the two dimensional world
of matrices, planar graphs and strings. Ordinary double scaling 
in tensor models \cite{doublescaling} does not lead out of the branched polymer phase.
Of course more interesting geometric phases may hide
in more sophisticated multiple-scaling limits of tensor models  or as
non-trivial fixed points of the renormalization 
group for tensor field theories \cite{tfts}, right now an active research field \cite{rgtensors}.

A second independent discovery
unexpectedly boosted the excitement about melons, namely the uncovering of the holographic and
maximally chaotic properties of the 
SYK model \cite{SYK}. These properties point to an 
interesting gravitational dual, and launched a new avenue of research. 
However the relationship between quantum gravity
and such a simple one dimensional model of condensed matter remains somewhat mysterious. 
Part of the veil was lifted when the tensor and SYK  research lines were related 
by tensor models \`a la Gurau-Witten \cite{GW}
which have the same chaotic properties than SYK but are \emph{bona fide} quantum theories. They are now seriously considered as providing the first computable toy models of \emph{truly quantum} black-holes \cite{SYKblackholes}. 
The study of tensor models on higher dimensional ordinary spaces 
and of their possible gravitational dual is also active and promising (e.g. \cite{higherdimtensors}).

Nevertheless it remains an open problem to connect these developments to
the initial random geometric motivation of tensor models \cite{Delporte:2018iyf}. 
This connection could happen 
through the development of a new ``random holography"
chapter of the gauge-gravity and AdS/CFT ongoing saga \cite{AdS}.

The present paper is a modest step in this direction.
We propose a systematic study of QFT on random trees, since we feel
that random trees form the most natural way to 
randomize the fixed time of SYK-type
quantum models.  The \emph{spine} common to all the infinite trees in the random 
sum allows to define convenient one dimensional 
observables for which the translation invariance and Fourier analysis of the SYK-type 
models remains available. This reminds of 
the two dimensional case, the Galton-Watson trees 
along the spine being a one-dimensional analog of the bumps of the Liouville field.
We would like to summarize this analogy in the bold statement
that pure quantum gravity in dimension 1 or ``gravitational time"
is simply ordinary time dressed by random lateral trees. 

We shall limit ourselves in this paper to perturbative results on Feynman amplitudes
for a self-interacting scalar theory. 
We take as propagator a fractional rescaled Laplacien as in \cite{Abd} to put ourselves in the interesting just renormalizable case. Our basic tool is
the multiscale analysis of Feynman amplitudes \cite{FMRS}, which remains available on random trees since it simply slices the proper time\footnote{This proper time is nothing but Feynman's parameter 
in high energy physics language.}  of the random path
representation of the inverse of the Laplacian.


Combining this slicing with the probabilistic estimates of Barlow and Kumagai \cite{BarlowKumagai,Kumagai}
we establish basic theorems on power counting, convergence and renormalization
of Feynman amplitudes. Our main results, Theorems \ref{theoconv} and \ref{theoconv1} below, use the Barlow-Kumagai technique of ``$\lambda$-good balls" to \emph{prove} that the  averaged amplitude of \emph{any} graph without superficially divergent subgraphs is finite and that logarithmically divergent graphs and subgraphs can be renormalized via local counterterms. We think that these
results validate the intuition that from the physics perspective
random trees indeed behave as an effective 
space of dimension 4/3. We postpone the more complete 
analysis of specific models and of their renormalization group flows and non-perturbative or constructive properties to the future. 

The paper is structured as follows. In section \ref{section:QFTonGraph}, we introduce the ensemble of random trees that will be of concern as well as the random walk approach to the propagator of the theory. We also recall the multiscale point of view for renormalization towards an infrared fixed point and motivate the rescaling of the Laplacian appropriate for just renormalizable models. After presenting briefly in section \ref{section:proba} the needed results of \cite{BarlowKumagai}, we prove upper and lower bounds on completely convergent graphs. In section \ref{section:localization}, we obtain upper bounds on differences on amplitudes when transporting external legs, important in order to assure local counterterms. Finally, we discuss in section \ref{section:comments} the setting that we think would stand for an analog of finite temperature field theory in this framework and the description of a model that would naturally serve as a concrete playground for the methods exposed below.

\section{Quantum Field Theory on a Graph}
\label{section:QFTonGraph}


\subsection{$\phi^{q}$ QFT on a graph}

For this introductory section we follow \cite{Gurau:2014vwa} (in particular its section 3.3.2).
Let us consider a space-time which is a proper  \emph{connected} graph $\Gamma$, 
with vertex set $V_\Gamma $ and edge set
$E_\Gamma$. It can be taken finite or infinite. The word ``proper" means that the graph has neither multiedges nor self-loops (often called tadpoles in physics). In the finite case we  often omit to write cardinal symbols such as
$\vert V_\Gamma \vert $, $\vert E_\Gamma \vert $ when there is no ambiguity.
In practice in this paper we shall consider mostly trees, more precisely 
either finite trees $\Gamma$ for which $V_\Gamma = E_\Gamma +1$, or infinite trees
in the sense of \cite{DJW}
which can be also interpreted as conditioned percolation clusters or Galton-Watson trees conditioned on non-extinction
in the sense of  \cite{BarlowKumagai}. The main characteristic of such infinite trees is to have a single 
\emph{infinite spine} $\cS(\Gamma) \subset V(\Gamma)$. This spine is decorated all along by lateral  independent  Galton-Watson finite critical trees, which we call the  branches,
see Figure \ref{treeandspine}.

On any such graph $\Gamma$, there is a natural notion of the Laplace operator $\cL_\Gamma$.
We recall that on a directed graph $\Gamma$
the \emph{incidence matrix}  is the rectangular $V$ by $E$ matrix 
with indices running over vertices and
edges respectively, such that 
\begin{itemize}
\item
${\epsilon_\Gamma(v,e)}$ is +1 
if $e$  ends at $v$, 
\item
${\epsilon_\Gamma(v,e)}$ is -1 if $e$ starts at $v$,
\item
${\epsilon_\Gamma(v,e)}$ is  0 otherwise.
\end{itemize}

The $V $ by $V$ square matrix 
with entries  $d_v$ on the diagonal is called the \emph{degree} or coordination  matrix
$D_\Gamma$. 
The {\emph{adjacency matrix}} is the symmetric $V \times V$ matrix $A_\Gamma$ made of zeroes on the diagonal: $A_\Gamma(v,v) = 0 \;\; \forall v\in V$, and such that 
if $v \ne w$ then $A_\Gamma(v,w)$  is the number of edges of $G$ which have vertices $v$ and $w$ as their ends. 
Finally the {\emph{Laplacian matrix}} of $\Gamma$ is defined to be $\cL_\Gamma = D_\Gamma - A_\Gamma$.
Its positivity properties stem from the important fact that it is a kind of square of the incidence matrix,
namely  
\bee
L_\Gamma = \epsilon_\Gamma \cdot \epsilon_\Gamma^\star. \label{laplacian}
\ee
Remark that this Laplacian is a positive rather than a negative operator (the sign convention being opposite to the one of differential geometry). Its kernel (the constant functions) has dimension 1 since $\Gamma$ is connected.

The kernel $C_\Gamma (x, y) $ of the inverse  of this operator  is formally given by the sum over random paths $\omega$ from $x$
to $y$
\beq  \cL_\Gamma^{-1} = C_\Gamma (x, y) =\left[\sum_n \left(\frac{1}{D_\Gamma}A_\Gamma\right)^n  \frac{1}{D_\Gamma} \right](x,y)= \sum_{\omega: x \to y} \; \prod_{v\in \Gamma} \;\biggl[\frac{1}{d_v}\biggr]^{n_v (\omega)}
\label{pathrep}
\eeq
where $d_v= D_\Gamma(v,v) = \cL_\Gamma (v,v)$ is the coordination at $v$ and $n_v (\omega)$ is the number of visits of $\omega$ at $v$. We sometimes omit the index $\Gamma$ when there is no ambiguity.

As we know this series is not convergent without an infrared regulator (this is related to the Laplacian having
a constant zero mode). 
For a finite $\Gamma$ we can take out this zero mode by fixing a root vertex in the graph and deleting 
the corresponding line and column in $\cL_\Gamma$.
But it is more symmetric to use the mass regularization. It 
adds $m^2 \bbbone$ to the Laplacian, where $\bbbone$ is the identity operator on $\Gamma $, with kernel $\delta (x,y)$. Defining $C_\Gamma^m (x, y)$ as the kernel of $(\cL_\Gamma + m^2 \bbbone )^{-1}$
we have the \emph{convergent} path representation
\bee  C_\Gamma^m (x, y) = \sum_{\omega: x \to y} \;  \prod_{v\in \Gamma} \;\biggl[\frac{1}{d_v + m^2}\biggr]^{n_v (\omega)}
\label{masslap}
\ee
and the infrared limit corresponds to $m \to 0$.

A scalar Bosonic free field theory $\phi$ on $\Gamma$ 
is a function $\phi : V_\Gamma \to \R$ defined on the vertices of the graph and
measured with the Gaussian measure 
\bee  d\mu_{C_\Gamma}(\phi)  =  \frac{1}{Z_0} e^{-\frac{1}{2} \phi (\cL_\Gamma + m^2 \bbbone ) \phi  
} \prod_{x\in V_\Gamma} d\phi(x) , \label{eq:mesuremu}
\ee
where $Z_0$ a normalization constant.
It is obviously well-defined as a finite dimensional probability measure for $\mu >0$ and $\Gamma$ finite.  
We meet associated infrared  divergences in the limit of $\mu=0$ and they are 
governing the large distance behavior of the QFT in the limit of infinite graphs $\Gamma$. The systematic
way to study QFT divergences is through a multiscale expansion 
in the spirit of \cite{FMRS,Rivabook,books,ChandraHairer}. No matter whether an ultraviolet or an infrared limit is considered, the renormalization group always flows from ultraviolet to infrared and the same techniques apply in both cases.

The $\phi^{q}$ interacting theory is then defined by the formal functional integral \cite{Gurau:2014vwa}:
\bee
d\nu_\Gamma (\phi )  =  \frac{1}{Z(\Gamma, \lambda)}  e^{- \lambda \sum_{x \in V_\Gamma} \phi^{q} (x) }  d \mu_{C_\Gamma} (\phi) \;,
\ee
where the new normalization is
\bee Z(\Gamma, \lambda)= \int e^{- \lambda \sum_{x \in V_\Gamma} \phi^{4} (x) }  d \mu_{C_\Gamma} (\phi) = \int d\nu_\Gamma (\phi ) .
\label{mesurenu}\ee

The correlations (Schwinger functions) of the $\phi^{4}$ model on  $\Gamma$ are
the normalized moments of this measure:
\bee  S_{N} (z_1,...,z_N )  = \int     
\phi(z_1)...\phi(z_N) \; d\nu_\Gamma (\phi ) , 
\ee
where the $z_i$ are external positions hence fixed vertices of $\Gamma$.
The case of fixed flat $d$-dimensional lattice corresponds to $\Gamma ={\mathbb Z}^d$.
As well known the Schwinger functions expand in the formal series of Feynman graphs
\bee  S_{N} (z_1,...,z_N )  = \sum_{V=0}^\infty \frac{(- \lambda)^V}{V!} \sum_G A_G (z_1,...,z_N ), 
\ee
where the sum over $G$ runs over Feynman graphs with $n$ internal vertices of valence $q$ and $N$ external leaves of valence 1.
Beware not to confuse these Feynman graphs with the ``space-time" 
graph $\Gamma$ on which the QFT lives.
More precisely $E_G$ is the disjoint union of a set $I_G$ of internal edges and of a set 
$N_G$ of external edges, and for the interaction $\phi^{q}$ these Feynman graphs have $V_G =V$ internal 
vertices which are regular with total degree $q$ and $N_G=N$ external leaves of degree 1. Hence $qV_G= 2E_G+N_G$. If $q$ is even this as usual implies parity
rules, namely $N_G$ has also to be even. We often write simply $V$, $E$, $N$ instead of $V_G$, $E_G$ and $N_G$
when there is no ambiguity.
In this paper, we do not care about exact combinatoric factors nor about convergence of this series
although these are  of course important issues treated elsewhere \cite{Rivabook}. We also shall
consider only connected Feynman graphs $G$, which occur in the expansion of the connected Schwinger functions.

As usual the treatment of external edges is attached to a choice for the external arguments of the graph.
Our typical choice here is to use external edges which all link a $q$-regular internal vertex to a 1-regular leaf with 
fixed external positions $z_1$, \dots $z_N$ in $\Gamma$. The (unamputated) graph amplitude is then a function of the external arguments obtained 
by integrating all positions $x_v$ of internal vertices $v$ of $G$ over our space time, which is $V(\Gamma)$. Hence
\bea   A_{G}(z_1, \cdots , z_N) &=&  \prod_{v\in V_G}\sum_{x_v \in V_\Gamma}  \prod_{\ell \in
E_G } C_\Gamma^m (x_{\ell},y_{\ell}) 
\eea
where $x_\ell$ and $y_\ell$ is our (sloppy, but compact!) notation for the vertex-positions at the two ends of edge $\ell$. 

We consider now perturbative QFT on \emph{random trees}, which instead of $\Gamma$ we note from now as $T$. 
The universality class of random trees \cite{aldous}
 is the Gromov-Hausdorff limit of any critical Galton-Watson tree process with fixed branching rate \cite{harris},
and conditioned on non-extinction.
 It has a unique \emph{infinite spine}, decorated with a product of independent Galton-Watson
measures for the branches along the spine \cite{DJW}. We briefly recall the corresponding probability measure,
following closely \cite{DJW}, but instead of half-infinite rooted trees with spine labeled by ${\mathbb N}$ we consider
trees with a spine infinite in both directions, hence labeled by ${\mathbb Z}$.

The order $\vert T \vert$ of a rooted tree is defined as its number of edges.
To a set of non-negative {\it branching weights} $w_i,\,i\in\mathbb
N^\star$ is associated the weights generating function $g(z):= \sum_{i \ge 1}  w_i z^{i-1} $ and the
{\it finite volume partition function} 
$Z_n$ on the set $\cT_n$ of all rooted trees $T$ of order $\vert T \vert =n$
\beq
Z_n = \sum_{T \in \cT_n}\prod_{u\in T \setminus r}w_{d_u}\,,
\eeq
where  $d_u$ denotes the degree of the vertex $u$.  The generating function for all $Z_n$'s
is 
\beq Z(\zeta ) = \sum_{n=1}^\infty Z_n \zeta^n  .
\eeq It satisfies the equation
\beq\label{fixZ}
Z(\zeta ) =\zeta g(Z (\zeta)).
\eeq
Assuming a finite radius of convergence $\zeta_0$ for $Z$ one defines 
\beq
Z_0 = \lim_{\zeta\uparrow \zeta_0}Z(\zeta ).
\eeq
The critical Galton-Watson
probabilities $p_i := \zeta _0 w_{i+1}Z_0^{i-1}$ for $i\in\mathbb N$ are then normalized: $\sum_{i=0}^\infty p_i = 1$. 
We then consider the class of infinite random trees defined by an \emph{infinite} spine of vertices $s_k$, $k\in \mathbb Z$,
plus a collection of $d_k -2$ \emph{finite} branches $T^{(1)}_k, \dots , T^{(d_k -2)}_k$, at each vertex $s_k$
of the spine (recall the degree of $k$ is indeed $d_k$). 
The set of such infinite trees is called $\cT_\infty$. It is equipped with a probability measure $\nu$ that we now describe.
This measure is obtained as a limit of measures $\nu_n$ on finite trees of order $n$. These measures $\nu_n$
are defined by \emph{identically and independently} distributing branches around a spine with measures 
\beq  \mu ( T) = Z_0^{-1}\zeta_0^{|T|}\prod_{u \in T \setminus r}w_{d_u}  = \prod_{i\in T\setminus r}p_{d_u -1}  \,.
\eeq

\noindent{\bf Theorem } \cite{DJW}
Viewing $\nu_n(T) = Z_n^{-1}\prod_{u\in T \setminus r}w_{d_u}\,,\quad\tau\in\cT_n\,,$ as a probability measure on $\cT$ we have
\beq
\nu_n \to \nu\quad \text{as}\quad n\to\infty\,,
\eeq
where $\nu$ is the probability measure on $\cT$ concentrated on the
subset of infinite trees $\cT_\infty$. 
Moreover the spectral dimension of generic infinite tree ensembles is
$ d_{spec} = 4/3\,$.

From now on we write $\mE (f)$ for the average according to the measure
$d \nu$ of a function $f$ depending on the tree $T$, and $\mP$ for the probability of an event $A$ according to $d\nu$. Hence $\mP (A) = \mE ( \chi_A)$ where $\chi_A$ is the characteristic function for the event $A$
to occur.
For simplicity and in order not to loose the reader's attention into unessential details we shall also
restrict ourselves from now on to the case of critical \emph{binary} Galton-Watson trees. 
It corresponds to weights $w_1 = w_3 = 1$, and $w_i = 0$ for all other values of $i$. 
In this case the above formulas simplify. 
The critical Galton-Watson process corresponds to offspring probabilities 
$p_0 =  p_2 = \frac{1}{2} $,  $p_i = 0$ for $i \ne 0, 2$.
The generating function for the 
branching weights is simply $g(z) = 1 + z^2$ and the generating function for the
finite volume trees $Z(\zeta ) = \sum_{n=1}^\infty Z_n \zeta^n  $
obeys the simple equation $
Z(\zeta ) =\zeta (1 + Z^2 (\zeta))$,
which solves to the Catalan function $Z= \frac{1 - \sqrt{1 - 4 \zeta^2}}{2\zeta}$. In the above notations 
the radius of convergence of this function is $\zeta_0 = \frac{1}{2}$. Moreover $Z_0 = \lim_{\zeta\uparrow \zeta_0}Z(\zeta ) =1$ and the independent measure on each branch of our random trees is simply
\beq
\mu (T) = 2^{- \vert T \vert }\,.
\eeq

\subsection{Fractional Laplacians}

Since the most interesting QFTs (including, in dimension 1, the tensorial theories \`a la Gurau-Witten) are the ones with just renormalizable power counting, 
we want to state our result in that case. A time-honored method  for that is to raise the ordinary Laplacian to a suitable fractional power $\alpha$  
in the QFT propagator \cite{Abd}. We assume from now on that this fractional power obeys  $0 < \alpha <1$ and call $C^\alpha$
 the corresponding propagator, i.e. the kernel of  $\cL^{-\alpha}$. It is most conveniently computed using the identity
\bee \cL^{-\alpha}  = \frac{\sin\pi\alpha}{\pi} \int_0^\infty \frac{2m^{1-2\alpha }}{\cL + m^2} dm \label{kallen}
\ee
since this ``K\"allen-Lehmann" representation respects the positivity
properties of the random path representation of the ordinary Laplacian inverse.

In the continuum $\R^d$ case, we have the ordinary heat kernel integral representation
\bea   
C^\alpha_{{\mathbb R}^d}(x,y) &=&  \frac{\sin\pi\alpha}{\pi} \int_0^\infty 2m^{1-2\alpha }dm \int_0^\infty e^{-m^{2}t -  
\frac{\vert x-y \vert^{2} }{4t }}  \frac{d t}{t^{d/2}} ,
\eea

On ${\mathbb Z}^d$ the kernel of the Laplacian between points
$x$ and $y$ is a slightly more complicated integral
\bee
C_{{\mathbb Z}^d}(x,y) =  \int_{0}^{2\pi} dp_1 \cdots \int_{0}^{2\pi} dp_d 
e^{i p \cdot (x-y)}    \left[2d -2 \sum_{i=1}^d \cos p_i\right]^{-1}.
\ee
but the kernel of the fractional Laplacian can be still written as an integral, namely
\bee
C_{{\mathbb Z}^d}^\alpha(x,y) =  \int_{0}^{2\pi} dp_1 \cdots \int_{0}^{2\pi} dp_d 
e^{i p \cdot (x-y)}    \left[2d -2 \sum_{i=1}^d \cos p_i\right]^{-\alpha}.
\ee
Hence using \eqref{kallen} it can also be represented as
\bea
C_{{\mathbb Z}^d}^{\alpha}(x,y) &=& \frac{\sin\pi\alpha}{\pi} \int_{0}^{2\pi} dp_1 \cdots \int_{0}^{2\pi} dp_d 
e^{i p \cdot (x-y)}\nonumber \\
&& \int_0^\infty \frac{2m^{1-2\alpha } dm}{m^2 + 2d -2 \sum_{i=1}^d \cos p_i} ,
\eea
%
leading to the random walk representation
\bee C_{{\mathbb Z}^d}^{\alpha}(x,y) = \frac{\sin\pi\alpha}{\pi}  \int_0^\infty 2m^{1-2\alpha } dm\sum_{\omega: x \to y}\prod_v \;\biggl[\frac{1}{2d + m^2}\biggr]^{n_v (\omega)} 
\ee
where $n_v (\omega)$ is the number of visits of $\omega$ at $v$.

As remarked above in the case of a general graph $\Gamma$ we no longer have translation invariance of Fourier integrals but still the 
random path expansion, so that
\bee C_{\Gamma}^{\alpha}(x,y) = \frac{\sin\pi\alpha}{\pi}  \int_0^\infty 2m^{1-2\alpha }dm\sum_{\omega: x \to y}\prod_v \;\biggl[\frac{1}{2d_v + m^2}\biggr]^{n_v (\omega)} 
\ee
where the walks $\omega$ now live on $\Gamma$ and $d_v$ is the degree at vertex $v$. 

\subsection{The Random Tree Critical Power $\alpha = \frac{2}{3} - \frac{4}{3q}$}

In integer dimension $d$, standard QFT \emph{power counting} with propagator $C^\alpha$ relies on 
the standard notion of degree of divergence. For  a regular Feynman graph of degree $q$ with $N$ external legs, this degree is defined as
\bee \omega(G) = (d-2 \alpha )E - d (V-1) = (d-2 \alpha )(qV-N)/2 - d (V-1) . \label{divdegree}
\ee

This power counting is neutral (hence does not depend on $V)$ in the critical or just-renormalizable case
\bee \alpha =\frac{ (q-2)d }{2q }  
\ee
in which case we have
\bee \omega(G) =d \left(1- \frac{N}{q}\right). \label{divdegree2}
\ee

For instance if $q=d=4$ we recover that the $\phi^4_4$ theory with propagator $p^{-2}$ is critical,
and if $d=1$ we recover the critical index 
$\alpha= \frac{1}{2} - \frac{1}{q}$ of the infrared SYK theory with $q$ interacting fermions \cite{SYK}. 

It turns out that the critical fractional power $\alpha$ which makes the probabilistic power counting of the $\phi^q$ QFT on random trees
just renormalizable, and the corresponding divergence degrees $\omega$ are  precisely obtained by substituting in these formulas the \emph{spectral dimension} $d=4/3$
of random trees, namely
\bee\alpha = \frac{2}{3} - \frac{4}{3q}\;, \quad \omega(G) =\frac{4-N}{3} . \label{crtiticalpha}
\ee  This is not surprising since this spectral dimension is precisely related to the short-distance,  long-time behavior of the inverse Laplacian
averaged on the random tree. We shall fix from now the fractional power $\alpha$ to its critical value and write $C_T$ instead of $C_T^\alpha$.
Nevertheless this simple rule requires justification, which is precisely provided 
 by the next sections.

\subsection{Slicing into Scales}

The multiscale decomposition of Feynman amplitudes is a systematic tool
to establish power counting and study perturbative and constructive renormalization in quantum field theory
\cite{Gurau:2014vwa,FMRS,Rivabook}. It relies on a sharp slicing into a geometrically growing sequence of scales of the
Feynman parameter for the propagator of the theory. This parameter is nothing but the \emph{time}
in the random path representation of the Laplacian. The short time behavior of the propagator is unimportant 
since the graph $\Gamma$ is an ultraviolet regulator in itself. We are therefore interested in infrared problems, namely the long distance behavior of the theory
(in terms of the graph distance). In the usual discrete
random walk expansion of the inverse Laplacian, the total time is the length of the path hence an 
integer. This integer when non trivial cannot be smaller than 1. However the results of \cite{BarlowKumagai}
are formulated in terms of a continuous-time random walk
which should have equivalent infrared properties. In what follows we shall use 
both points of view.

\begin{definition}[Time-of-the-Path Slicing]
The infrared parametric slicing of the propagator is
\bea  C &=&  \sum_{j=0}^{\infty}  C^j  : \quad C^0= \bbbone,  \nonumber\\
C^j &=&  \sum_{\omega: x \to y \atop M^{2(j-1)} \le n(\omega) <  M^{2j} } 
\;  \prod_{v\in \Gamma} \;\biggl[\frac{1}{d_v + m^2}\biggr]^{n_v (\omega)}  \quad \forall j \ge 1. \label{decoa}
\eea
\end{definition}
$M$ is a fixed constant which parametrizes the thickness of a renormalization group slice 
(the craftsman trademark of \cite{Rivabook}).
Each propagator  $C^{j}$ indeed corresponds to a theory with
both an ultraviolet and an infrared cutoff, which differ by the fixed 
multiplicative constant $M^2$. 
An infrared cutoff on the theory is then obtained by setting a maximal value $\rho= j_{max}$ for the  index $j$.
The covariance with this cutoff is therefore 
\bee  C_{\rho} =  \sum_{j=0}^{\rho}  C^{j} .  \label{decocut1} 
\ee
In the continuum $\R^d$ case we have the ordinary heat kernel representation
hence the explicit integral representation
\bea   
C^{j}_{{\mathbb R}^d}(x,y) &=&  \frac{\sin\pi\alpha}{\pi} \int_0^\infty 2m^{1-2\alpha }dm  \int_{M^{2j}}^{M^{2(j+1)}} e^{-m^{2}t -  
\frac{\vert x-y \vert^{2} }{4t }}  \frac{dt}{t^{d/2}} 
\eea
from which it is standard to deduce scaling bounds such as 
\bea C_{{\mathbb R}^d}^{j}(x,y) &\le& K M^{(2 \alpha -d)j} e^{- c M^{-2j}\vert x -y \vert^2} . \label{flatbound}
\eea
for some constants $K$ and $c$. From now on in this paper we use most of the time $c$ as a generic name for
any inessential constant ($c$ is therefore the same as the $O(1)$ notation 
in the constructive field theory literature). 
We shall also omit from now on to keep
inessential constant factors such as $ \frac{\sin\pi\alpha}{\pi} $. 
In ${\mathbb Z}^d$ the sliced propagator then writes
\bee C_{{\mathbb Z}^d}^{j}(x,y) = \int_0^\infty 2m^{1-2\alpha } dm \sum_{\omega: x \to y \atop M^{2(j-1)} \le n(\omega) <  M^{2j} } \prod_v \;\biggl[\frac{1}{2d + m^2}\biggr]^{n_v (\omega)} .
\ee
It still can be shown easily to obey the same bound  \eqref{flatbound}. For
a general tree $T$ the sliced decomposition of the propagator then writes
\bea  C_{T}(x,y) &=&  \sum_{j=0}^{\infty}  C_{T}^{j}(x,y)  : \quad C_{T}^{0}= \bbbone,  \quad {\rm and\;\; for }\;\;  j \ge 1, \label{slicedprop1}  \label{decoa1} \\
C_{T}^{j}(x,y)  &=&   \int_0^\infty 2m^{1-2\alpha } dm \sum_{\omega: x \to y \atop M^{2(j-1)} \le n(\omega) <  M^{2j} } 
\;  \prod_{v\in T} \;\biggl[\frac{1}{d_v + m^2}\biggr]^{n_v (\omega)}.
\label{slicedprop2}
\eea
Remark that after $n$ steps a path cannot reach farther than distance $n$ (for the discrete time random walk). In particular we can safely include the
function $\chi_j (x,y)$ in any estimate on $C_{T}^{j}$, where $\chi_j (x,y)$ is the characteristic function for
$d(x,y) \le M^{2j}$\footnote{$d(x,y)$ denotes the smallest number of steps on the tree needed to connect $x$ to $y$.}.
A generic tree $T$ in $\cT$ has spectral dimension $4/3$ so that we should expect for such a tree 
\bea C_{T}^{j}(x,y) &\le& K M^{(2 \alpha -\frac{4}{3})j} \chi_j (x,y) . \label{flatbound1}
\eea
A fixed tree however can be non-generic, hence has no a priori well defined dimension $d$. Since it nevertheless
always contains an infinite spine which has dimension 1, we expect that for any tree $T$ in $\cT$
the following bound holds
\bea C_{T}^{j}(x,y) &\le& K M^{(2 \alpha -1)j}\chi_j (x,y) . \label{flatbound2}
\eea
However we do not need a very precise bound for exceptional trees since as we will see in the next section, they will 
be wiped by small probabilistic factor. 
In fact a very rough ``dimension zero" bound can be obtained easily for all points $x$, $y$ on $T$:
\bea  C_{T}^{j}(x,y)  &\le& K M^{2 \alpha j} \chi_j (x,y). \label{flatbound3}
\eea
Indeed, this bound is easy to obtain by simply overcounting the number of paths from $x$ to $y$ in time $t$ as the total number of paths from $x$ in time $t$. 
In the binary tree case each vertex degree is bounded by 3. At a visited vertex $v$ we have $d_v$ choices for the next random path step so that
\bea\sum_{\omega: x \to y \atop M^{2(j-1)} \le n(\omega) <  M^{2j} } \;  \prod_{v\in T} \;\biggl[\frac{1}{d_v + m^2}\biggr]^{n_v (\omega)} &\le& 
\sum_{M^{2(j-1)} \le n<  M^{2j} } \;  \left[\frac{3}{3+m^2}\right]^{n}\\
&\le& K \int_{M^{2(j-1)}}^{M^{2j}} dt e^{-ctm^2}
\eea
where $K$ and $c$
are some inessential constants. Then the naive inequality
\bee K \int_0^\infty m^{1-2 \alpha}dm 
\int_{M^{2(j-1)}}^{M^{2j}} dt e^{-ctm^2} \le K' M^{2j\alpha},
\ee
allows to conclude.

However none of the bounds \eqref{flatbound1}-\eqref{flatbound3} are sufficient to establish the correct
power counting of Feynman amplitudes averaged on $T \in \cT$. We need to combine the multiscale decomposition
(best tool to estimate general Feynman amplitudes on a fixed space) with probabilistic estimates 
to show that the prefactor $ M^{(2 \alpha -\frac{4}{3})j} $ in \eqref{flatbound1}
is indeed the typical one and that the typical volume factors for the integrals on vertex positions
correspond also to those of a space of dimension 4/3.

\subsection{The  Multiscale Analysis}

Consider a fixed connected Feynman graph $G$ with $n$ internal 
 vertices, all with degrees $q=4$, $N$ external edges and $L =2n -N/2$ internal edges. 
There are in fact several possible prescriptions to treat external arguments in a Feynman amplitude \cite{FMRS,Rivabook},
but they are essentially equivalent from the point of view of integrating over inner vertices the product of propagators. 
A convenient and simple choice is to put all external legs in the most infrared scale, namely the infrared cutoff scale $\rho$
(similar to a zero external momenta prescription in a massive theory),
and to work with amputated amplitudes which no longer depend
on the external positions $z_1, \dots, z_N$ but only of the position 
$x_0$ of a fixed inner \emph{root vertex} $v_0$. It means we 
forget the  $N(G)$ external propagators $C_{T}(x_{v(k)},z_k) $ factors in $A_G$ and shall integrate only the $n-1$ positions 
$x_v, v \in \{1, \dots, n-1 \}$. In this way we get an amplitude
$A_G^{amp} (x_0)$ which is solely a function\footnote{In a usual  theory there is no $x_0$ dependence because of translation invariance, but for a particular tree $T$ there is no such invariance.} of $x_0$.
However we should remember that fields and 
propagators at the external cutoff scale have a canonical  dimension
which in our case for a field of scale $j$ is $M^{-j/3}$. To compensate 
for the missing factors after amputation
we shall multiply this 
amputated amplitude by $M^{-\rho N /3}$, and for the fixing of 
position $x_0$, we shall add another global factor $M^{4\rho/3}$. 
Hence we define
\bee \tilde A_G^{amp} (x_0) :=  M^{\rho (4-N) /3}
\prod_{v=1}^{n-1} \sum_{x_v \in V(T)} \prod_{\ell \in I(G) } C_{T} (x_\ell, y_\ell ). \label{agmu0}
\ee
For simplicity, we write now $A_G$ again, instead of $\tilde A^{amp}_G$.
The decomposition \eqref{decoa1} leads to the multiscale representation for a Feynman 
graph $G$, which is:
\bea   A_{G}(x_0) &=&  M^{\rho (4-N) /3} \sum_{\mu} A_{G,\mu}(x_0)\;  \; ,\label{agmu1}\\
A_{G,\mu }(x_0) & =& \prod_{v=1}^{n-1} \sum_{x_v \in V(T)}  
\prod_{\ell \in I(G) }   C_{T}^{j_\ell} (x_\ell, y_\ell ) . \label{agmu2}
\eea
$\mu$ is called a ``scale assignment" (or simply ``assignment").
It is a list of integers $\{ j_\ell \}$, one for each internal edge of $G$, which  provides for each internal edge $l$ of $G$ the scale $j_\ell$ of that edge.  $A_{G,\mu}$ is the amplitude associated to the pair $(G,\mu)$, and \eqref{agmu1}-\eqref{agmu2} is the multiscale representation of the Feynman amplitude.

We recall that the key notion in the multiscale analysis of a Feynman amplitude is that of ``high" subgraphs. In our infrared setting, this means the connected components of $G_{j}$, the subgraph of $G$ made of all edges $\ell$ with index $j_\ell \le j$. These connected components  are labeled as $G_{j,k}$, $k=1,...,k(G_j)$, where $k(G_j)$ denotes the number of connected components of the graph $G_j$.

A subgraph $g \subset G$ then has  in 
the assignment $\mu$ internal and external indices defined as
\bee   i_g(\mu)  = \sup\limits_{l {\rm \ internal \ edge \ of \ } g}  \mu(l)  \label{II.1.10}
\ee
\bee   e_g(\mu)  = \inf\limits_{l {\rm \ external \ edge \ of \ } g}  
\mu(l) . \label{II.1.11}
\ee
Connected subgraphs verifying the condition 
\bee   e_g(\mu)  > i_g(\mu)  \quad  \quad {\rm (high \ condition)}  
\label{II.1.12}
\ee
are exactly the \emph{high} ones. This definition depends on the assignment $\mu$. For
a high subgraph $g$ and any value of $j$
such that $i_g (\mu )  < j \le e_g(\mu ) $ there exists exactly one value of
$k$ such that $g$ is equal to a $G_{j,k}$. 
High subgraphs are partially ordered by inclusion and form a forest in the sense of 
inclusion relations \cite{FMRS,Rivabook}.

The key estimates then keep only the spatial decay of a $\mu$-optimal spanning tree $\tau(\mu)$ of $G$,
which minimizes $\sum_{\ell \in \tau} j_\ell (\mu)$
(we use the notation $\tau$ for spanning trees of $G$
in order not to confuse them with the random tree $T$). The important 
property of $\tau (\mu)$ is that it is a spanning tree
within each high component $G_{j,k}$ \cite{FMRS,Rivabook}.
It always exists and can be chosen according to Kruskal greedy algorithm \cite{kruskal}.
It is unique if every edge is in a different slice; otherwise there may be several such trees in which case one simply picks one of them.

Suppose we could assume bounds similar to the flat case. It would mean that 
a sliced propagator in the slice $j_\ell$ would be bounded as
\bee  C_{T}^{j_\ell} (x_\ell, y_\ell ) \simeq K M^{-2j_\ell/3} e^{-M^{-j_\ell}  d(x_\ell,y_\ell) }
\ee
and that spatial integrals over each $x_v$ would be really 4/3 dimensional, i.e cost $M^{4j/3}$ if performed with the decay of a scale $j_v$ propagator.
Picking a Kruskal tree $\tau (\mu) $ with a fixed root vertex, and forgetting the spatial decay of all the edges not in $\tau$, one can then recursively organize
integration over the position $x_v$ of each internal vertex $v$ from the leaves towards the root. This can be indeed done using for each $v$ 
the spatial decay of the propagator joining $v$ to its unique towards-the-root-ancestor 
$a(v)$ in the  Kruskal tree. In this way calling $j_v$ the scale of that propagator 
we would get as in \cite{FMRS,Rivabook} an estimate
\bea
|A_{G,\mu}| &\le& K^{V(G)} M^{-N \rho/3}   \prod_{\ell \in I(G)} M^{-2j_\ell/3} \prod_{v \in V(G)} M^{4j_v/3} \\
&=& K^{V(G)} \prod_{j =1}^\rho \prod_{k=1}^{k(G_j)}   M^{\omega(G_{j,k})} \label{sketch}
\eea
where the divergence degree 
of a subgraph $S \subset G$ is defined as
\bee \omega(S) = \frac{2}{3}E(S) - \frac{4}{3} (V(S)-1) = \frac{4-N(S)}{3} .
\ee

Standard consequences of such bounds are 
\begin{itemize}
\item uniform exponential bounds for completely convergent graphs \cite{Rivabook}.
\item renormalization analysis: when high subgraphs have positive divergent degree 
we can efficiently replace them by local counterterms, which create a flow for marginal and relevant operators.
The differences are remainder terms which become
convergent and obey the same bounds as for convergent graphs, provided we use
an effective expansion which renormalizes only high subgraphs \cite{FMRS,Rivabook}.
\end{itemize}

In fact these bounds cannot be true for all particular trees $T$ since they depend on the Galton-Watson branches being typical. In more exceptional cases, for instance for a
tree reduced to the spine plus small lateral branches the effective spatial dimension is 1 rather than 4/3.
Such exceptional cases become more and more unlikely when we consider larger and larger sections of the spine. Our probabilistic analysis below proves that for the \emph{averaged} Feynman amplitudes everything happens as in equation \eqref{sketch}. To give a meaning to these averaged amplitudes, we fix the position of  the root vertex $x_0$ to lie on the spine of $T$. Averaging over $T$
restores translation invariance along the spine, so that we have finally to evaluate averaged amplitudes $\mE (A_G)$ which are simply numbers. It is for these amplitudes that 
we shall prove in the next sections our main results
Theorem \ref{theoconv} and \ref{theoconv1}.
But we need to introduce first our essential probabilistic tool, namely the $\lambda$-good conditions on trees of \cite{BarlowKumagai}.

\section{Probabilistic Estimates}
\def\lam {\lambda}
\label{section:proba}
We have first to recall the probabilistic estimates on random trees from
\cite{BarlowKumagai} that we are going to use, simplifying slightly some inessential aspects.
As mentioned above, \cite{BarlowKumagai} mostly considers random paths 
which are Markovian processes with \emph{continuous} times, but 
those are statistically equivalent to discrete paths in the interesting long-time infrared limit.

For $x \in T$, we note $B(x,r)$ the ball of $T$ centered on $x$ and containing points at most at distance $r$ from $x$,
and $m(x,r)$ the number of points of $T$ at distance $1+[r/4]$ of $x$, where $[.]$
means the integer part. For $(x,y) \in T^2$, we also write $q_t (x,y)$ (or sometimes $q_{t,x}(y)$) for the sum over random paths  in time $t$.

For   $\lam\ge 64$,
the ball $B(x,r)$ is said {\sl $\lam$--good} (Definition 2.11 of \cite{BarlowKumagai}) if:
\bea
r^2\lam^{-2} &\le& V(x,r) \le r^2\lam  , \label{goodvol} \\
m(x,r) &\le& \frac{1}{64} \lam, \quad
V(x, r/\lam) \ge r^2 \lam^{-4}, \quad V(x, r/\lam^2) \ge r^2 \lam^{-6}.
\eea
Remark that if $B(x,r)$ is {\sl $\lam$--good} for some $\lambda$,
it is {\sl $\lam'$--good} for all $\lambda' > \lambda$.

Corollary 2.12 of \cite{BarlowKumagai} proves that
\bee\mP  (B(x,r) \hbox{ is not $\lam$--good}) \le c_1 e^{-c_2 \lam}. 
\label{smallproba1}
\ee
This inequality together with the Borel-Cantelli lemma imply that given $r$ and a discrete monotonic sequence $\{\lam_{l}\}_{l\geq 0}$ with $\lim_{l \to \infty} \lam_{l}= +\infty $, there is, with probability one, a finite $l_0$ such that $B(x,r)$ is $\lam_{l_0}$-good  (see also the proof of Theorem 1.5 in \cite{BarlowKumagai}). In particular
\begin{lemma}\label{smallproba3}
Defining the random variable $L=\min \{l : B(x,r) \text{ is } \lam_{l}\text{-good}\}$
we have
\bee \mP [ L=l]
\le c_1 e^{-c_2 \lam_{l-1}}. \label{smallproba2}
\ee
\end{lemma}
\prf This is because the ball $B(x,r)$
must then be $\lam_{l-1}\text{-bad}$.
\qed

\medskip
Besides, the conditions of $\lambda$-goodness allow to bound with the right scaling the random path factor  $q_t (x,y)$ for $y$ not too far from $x$. More precisely the main part of Theorem 4.6 of \cite{BarlowKumagai} reads
\begin{theorem}
\label{theoBK1}
Suppose that $B=B(x,r)$ is $\lam$--good for 
$\lambda\ge 64$, and let  $I(\lam,r)=[r^3 \lam^{-6}, r^3 \lam^{-5}]$. Then 

\begin{itemize}
\item 
for any $K \ge 0$ and any $y\in T$ with $d(x,y) \le  Kt^{1/3}$
\bee  q_{2t}(x,y) \le  c\left(1+{\sqrt K}\right) t^{-2/3} \lam^{3}
\quad \hbox{\rm for } t\in I(\lambda, r) \;,  \label{gooddecay1}
\ee

\item
for any $y\in T$ with $d(x,y) \leq c_2 r \lambda^{-19}$
\bee  q_{2t}(x,y) \ge  c t^{-2/3} \lam^{-17}
\quad \hbox{\rm for } t\in I(\lambda, r).  \label{gooddecay2}
\ee

\end{itemize}

\end{theorem}
Notice that these bounds are given for for $q_{2t}(x,y)$ 
but the factor $2$ is inessential (it can be gained below
by using slightly different values for $K$) and we omit it from now on for simplicity. 

\subsection{Warm-up}

To translate these theorems into our multiscale setting, we introduce
the notation $I_j = [ M^{2(j-1)},  M^{2j}] $ and we have the infrared equivalent continuous time representation
\bee C^j_T (x,y) = \int_0^\infty u^{-\alpha } du \int_{I_j}  q_t (x,y) e^{-ut } dt = \Gamma(1- \alpha)\int_{I_j}  q_t (x,y)t^{\alpha -1} dt \label{trans}
\ee
which relates our sliced propagator 
\eqref{slicedprop2}
to the kernel $q_t$ of \cite{BarlowKumagai}. We forget from now on 
the inessential $\Gamma(1- \alpha)$
factor.
In our particular case $q=4,\alpha= 1/3$, \eqref{trans} means that  
we should simply multiply the estimates on $q_t$ established in \cite{BarlowKumagai} 
by $c M^{2j/3}$ to obtain similar estimates for $ C^j_T$. However we have also to perform
spatial integrations not considered in \cite{BarlowKumagai}, which complicate the probabilistic analysis.
As a warm up, let us therefore begin with a few very simple examples. Recall
that we do not carefully track inessential constant
factors in what follows, and that we can use the generic letter $c$ for any such constant
when it does not lead to confusion.

\begin{lemma} [Single Integral Upper Bound]\label{oneline}
There exists some constant $c$ such that
\bee  \mE \left[ \sum_y C_T^j (x,y) \right] \le c M^{2j/3} .
\label{eq:lemma-singleline}
\ee
\end{lemma}
\prf
We introduce two indices $k \in \N$, and $l \in \N$ with the condition $l\ge l_0 := \sup \{M^{2}, 64\}$ and parameters
$\lambda_{k,l}:= k+ l $. We also define radii
\bea r_{j,k}  &:=&  M^{2j/3} k^{5/3} ,\\ 
r_{j,k,l} &:=&  M^{2j/3}(k+ l)^{5/3},
\eea
and the balls $B^T_{j,k}$ and $B^T_{j,k,l}$ centered on $x$
with radius $r_{j,k}$ and $r_{j,k,l}$ (we put an upper index $T$
to remind the reader that these sets depend on our random space, 
namely the tree $T$). We also define the annuli
\bee
A^T_{j,k} := \{y : d(x,y) \in   [r_{j,k},r_{j,k+1}[
\},
\ee
so that the full tree is the union of the annuli $A^T_{j,k}$
for $k \in \N$:
\bee  T = \cup_{k \in \N} \; A^T_{j,k}. \label{sumT}
\ee
Remark that $A^T_{j,k} \subset B^T_{j,k+1} \subset B^T_{j,k,l}$ 
for any $l \in \N^\star$.
Remark also that with these definitions
\bee I_j = [M^{2j-2},M^{2j}] \subset I(\lambda_{k,l},r_{j,k,l})=
[r_{j,k,l}^3\lambda_{k,l}^{-6}, r_{j,k,l}^3\lambda_{k,l}^{-5}],
\ee
where $I(\lambda,r)$ is as in Theorem \ref{theoBK1},
since our condition $l \ge l_0 \ge M^{2}$ ensures that $r_{j,k,l}^3\lambda_{k,l}^{-6} \le M^{2j-2}$. 
Finally defining $K_k := M^{2/3} (k+1)$ we have
\bee  d(x,y ) \le K_k t^{1/3},\quad  \forall t \in I_j,\;\forall y \in A^T_{j,k}. \label{distcond}
\ee
Since the propagator is pointwise positive we can commute any sum or integral as desired. Taking \eqref{sumT} into account we can
organize the sum over $y$ according to the annuli $A^T_{j,k}$. Commuting the 
sum $\mE$ and the sum over $k$, 
according to the Borel-Cantelli argument in the section above, there exists (almost surely in $T$) a smallest finite $l$ such that the $B^T_{j,k,l}$ ball is $\lam_{k,l}$-good. Defining the random variable $L=\min \{l \ge l_0 : B^T_{j,k,l} \text{ is } \lam_{k,l}\text{-good}\}$, we can
partition our $\mE$ sum according to
the different events $L=l$. We now fix this $l$ so as to evaluate, according to \eqref{trans}
\bee \mE \left[\sum_y C_T^j (x,y) \right] = \sum_{k=0}^\infty \sum_{l=l_0}^\infty \mP[L=l] \mE\vert_{L=l} \Bigl[\sum_{y \in A^T_{jk}} \int_{I_j} dt t^{\alpha-1} q_t (x,y) 
\Bigr],
\ee
where $\mE\vert_{A}$ means conditional expectation with respect to the event $A$.
We are in a position to apply Theorem \ref{theoBK1} since all hypotheses and conditions are fulfilled (including 
$\lambda_{k,l} \ge 64$ since $l_0 \ge 64$). We have for some inessential constant $c$, under condition $L=l$
\bee  q_t (x,y)  \le c (1 + \sqrt{K_k} )
M^{-4j/3} \lam_{k,l}^3 , \quad  \forall t \in I_j,\;\forall y \in A^T_{j,k} .
\ee
Hence integrating over $t \in I_j$ 
\bee  C_T^j (x,y)  \le  c (k+l)^{7/2} M^{-2j/3},\quad  \forall y \in A^T_{j,k}, \label{cjesti1}
\ee
for some other inessential constant $c$.
We can now sum over $y \in A^T_{j,k}$, overestimating the volume of the annulus $A^T_{j,k}$ by the volume of the $B^T_{j,k,l}$ ball, to obtain
\bee \sum_{y \in A^T_{j,k}} C_T^j (x,y)  
\le  c  (k+l)^{7/2} M^{-2j/3} vol(B^T_{j,k,l}) .\label{cjesti12}
\ee
The condition $L=l$
allows to control the volume $vol(B^T_{j,k,l}) $ by the $\lam_{k,l}\text{-good}$ condition.
More precisely  \eqref{goodvol} implies
\bee
\mE\vert_{L=l} \, [ vol(B^T_{j,k,l}) ] \le  r_{j,k,l}^2 \lambda_{k,l} .
\ee
Using Lemma \ref{smallproba3} we conclude that
\begin{align}
\mE \left[\sum_y C_T^j (x,y) \right] &\le c\sum_{k=0}^\infty \sum_{l=l_0}^\infty \mP[L=l]
(k+l)^{7/2}M^{-2j/3}  r_{j,k,l}^2 \lam_{k,l}\nonumber
 \\
&\leq c  M^{2 j/3}\sum_{k=0}^\infty \sum_{l=l_0}^\infty  e^{-c'(k+l)} (k+l)^{47/6} \le c  M^{2 j/3} . 
\end{align}

\qed 
\begin{cor}[Tadpole]\label{tadpolecor}
  There exists some constant $c$ such that
\bee \mE \left[ C_T^j (x,x) \right] \le c M^{-2j/3} .
\label{cortadpole}
\ee
\end{cor}
\proof
We apply the same reasoning but instead of summing over $y$ we simply use \eqref{cjesti1} to get
\begin{align}
\mE \left[ C_T^j (x,x) \right] &= 
c\sum_{k=0}^\infty \sum_{l=l_0}^\infty \mP[N=l] M^{-2j/3 } (k+l)^{7/2} \nonumber
\\
&=c M^{-2j/3} \sum_{k=0}^\infty \sum_{l=l_0}^\infty  e^{-c'(k+l)}
(k+l)^{7/2} \le c  M^{-2 j/3},
\end{align}
since \eqref{distcond} is automatically satisfied for $y=x$.
\qed

\medskip
A lower bound of the same type is somewhat easier, as we do not need to exhaust the full spatial integral but can restrict to a subset, in fact a particular $\lambda$-good ball.
\begin{lemma}[Single Integral Lower Bound] \label{lowerbou}
\bee  \mE \left[ \sum_y C_T^j (x,y) \right] \ge c M^{2j/3} .\label{lowerbou1}
\ee
\end{lemma}

\prf 
We follow the same strategy than for the upper bound but we do not need
the index $k$ and the annuli $A_{j,k}$, since most of the volume is typically in the first annulus - 
namely the $k=0$  ball $B_j$. Restricting the sum over $y$  this ball is typically enough 
for a lower bound of the \eqref{lowerbou1} type. So we work at $k=0$ but we need again probabilistic estimates to tackle
the case of untypical volume of the ball $B_j$.
Therefore we define 
for $l\ge l_0 := \sup \{M^{2}, 64\}$, the parameter 
$\lambda_l = l$ and the two balls $B^T_{j,l} = B(x,r_{j,l})$ and 
$\tilde B^T_{j,l} = B(x,\tilde r_{j,l})\subset B^T_{j,l}$ of radii
respectively $r_{j,l}:=M^{2j/3}\lambda_l^{5/3}$ and $\tilde r_{j,l}:= c_2r_{j,l}\lambda_l^{-19}$ (in order for \eqref{gooddecay2} to apply below). We introduce the random variable 
\bee L=\min \{l \ge l_0 : B^T_{j,l} \text{ and }  \tilde B^T_{j,l}  \text{ are both } \lam_{l}\text{-good}\}.
\ee

Again, our choice of $r_{j,l}$ ensures that
\bee I_j = [M^{2j-2},M^{2j}] \subset I(\lambda_{l},r_{j,l})=
[r_{j,l}^3\lambda_{l}^{-6}, r_{j,l}^3\lambda_{l}^{-5}],
\ee
and the summands being positive, we will restrict the sum over $y$ to the smaller ball $\tilde B^T_{j,l} \subset B^T_{j,l}$, in order for \eqref{gooddecay2} to apply. We get
\bea 
\mE \left[\sum_y C_T^j (x,y) \right] 
&\ge& 
\mP[L\le l] \mE\vert_{L \le l} \
\Bigl[\sum_{y \in \tilde B^T_{j,l} } \int_{I_j} dt t^{\alpha-1} q_t (x,y)
\Bigr], \quad \forall l,\  \\
&\ge& c M^{-2j/3} l^{-17} \mP[L\le l] \;\mE\vert_{L=l} 
[ vol( \tilde B^T_{j,l} ) ],\quad\quad \forall l,\ \\
&\ge&  cM^{2j/3}\mP[L\le l] l^{-166/3}, \quad\quad\quad\quad\quad\quad\quad\quad\ \forall l, \  \\
&\ge&  cM^{2j/3} .
\eea
Indeed for the last inequality we remark that 
$\lim_{l \to \infty}\mP[L\le l] = 1$ (by Lemma \ref{smallproba3}) 
hence $\sup_{l \ge l_0}\mP[L\le l]l^{-166/3} $ is a \emph{strictly positive} constant that we absorb in $c$.
\qed 

\subsection{Bounds for Convergent Graphs}\label{fullgraphs}

In this section we prove our first main result, namely 
the convergence of Feynman amplitudes 
of the type \eqref{agmu0}-\eqref{agmu2} as the infrared cutoff $\rho$ is lifted.
Therefore we consider a fixed completely convergent graph $G$ with $n$ inner vertices and $N$ external lines, hence for which $N(S) \ge 6\; \forall S \subset G$. In this graph we mark a root vertex $v_0$ with fixed position $x_0$, 
lying on the spine, i.e. common to all trees $T$. By translation invariance of the infinite spine, the resulting amplitude $A_G (x_0)$
is in fact independent of $x_0$ and we have

\begin{theorem}\label{theoconv} For a completely convergent graph (i.e. with no 2 or 4 point subgraphs)
$G$ of order $V(G) = n$, the limit as 
$\lim_{\rho \to \infty} \mE ( A_G)$
of the averaged amplitude exists 
and obeys the uniform bound
\bee \mE ( A_G) \le K^n (n!)^\beta \label{uniconv}
\ee
where $\beta =\frac{52}{3}$. \footnote{We do not try to make $\beta$ optimal. We expect that a tighter probabilistic analysis could prove
subfactorial growth in $n$ for $\mE ( A_G)$.}
\end{theorem}

\proof
From the linear decomposition $A_{G} = \sum_\mu A_{G,\mu}$
follows that $\mE (A_G) = \sum_\mu \mE (A_{G,\mu} )$.
As mentioned above we use
only the decay of the propagators of an optimal Kruskal 
tree $\tau (\mu)$ to perform the spatial integrals over the position of the inner vertices. It means that we first 
apply a Cauchy-Schwarz inequality to  the $n+1 - N/2$
edges $\ell\not\in \tau (\mu)$. Labeling all the corresponding half-edges (not in $\tau(\mu)$)  as fields $f= 1, \cdots 2n+2 - N$ and their positions and scale as $x_f$ and $j_f$ we have
\bea \prod_{\ell \not \in \tau (\mu)} 
C_T^{j_\ell} (x_\ell,y_\ell) &\le& \prod_{\ell \not \in \tau (\mu)} 
\sqrt{C_T^{j_\ell}(x_\ell, x_\ell) C_T^{j_\ell}(y_\ell, y_\ell) }\nonumber
\\
&=&\prod_{f=1}^{2n+2 - N} [C_T^{j_f}(x_f,x_f)]^{1/2}.
\label{loopbou}
\eea

Each inner vertex $v\in \{1, \cdots n-1\}$ to integrate over is linked to the root by a single path in $\tau (\mu)$. The first line, $\ell_v$, in this path relates $v$ to a single ancestor $a(v)$ by an edge $\ell_v \in \tau (\mu)$. This defines a scale $j_v := j_{\ell_v}(\mu)$ for the sum over the position $x_v$.  

Taking \eqref{loopbou} into account, we write therefore 
\bee \mE [A_{G,\mu }] \le \mE \Big[  
\sum_{\{x_{v} \}} 
\prod_{v=1}^{n-1} C_{T}^{j_v} (x_v, x_{a(v)})
\prod_{f =1}^{2n+2 - N} [C_T^{j_f}(x_f,x_f)]^{1/2} \Big] \; . \label{agmu3}
\ee

We apply now to the $n-1$ spatial integrals exactly the same analysis
than for the single integral of Lemma \ref{oneline}.
The main new aspect is that the events of the previous section do not provide \emph{independent} small factors for each spatial integral. 
For instance if two positions $x_v$ and $x_{v'}$ 
happen to coincide and the smallest-$l$ $\lambda_{l}$-good 
event occur for a ball centered at $x_v$, it \emph{automatically implies}
the $\lambda_{l-1}$-bad event for the ball centered at $x_{v}$
and at $x_{v'}$,
because it is the \emph{same event}. Therefore in this case we do not get twice the same small associated probabilistic factor
of Lemma \ref{smallproba3}. This is why we loose a (presumably spurious)
factorial $[n!]^\beta$ in \eqref{uniconv}.

More precisely we introduce for each $v \in [1, n-1]$ two
integers $k_v$ and $l_v \ge l_0$, 
the radii  $r_{j_v, k_v}$, $r_{j_v, k_v, l_v}$
and the parameters $\lambda_{k_v,l_v}$ exactly as before.
 We introduce also
all these variables for every field $f \in [ 1, \cdots 2n+2 - N ]$ not in $\tau( \mu)$. We define again the random variable $L_v$ for $v \in [1, n-1]$ as the first integer $\ge l_0$ such that the ball 
$B^T_{j_v,k_v,l_v}$ is  $\lam_{k_v,l_v}\text{-good}$ and
$L_f$ for $f \in [1, n+1-N/2]$ as the first integer $\ge l_0$ such that the ball 
$B^T_{j_f,k_f,l_f}$ is  $\lam_{k_f,l_f}\text{-good}$.
The integrand is then bounded according to Theorem \ref{theoBK1},
leading to
\bea \mE [A_{G,\mu }] &\le& c^n
\sum_{\{k_v\},\{l_v\}\atop\{k_f\},\{l_f\}}\mP (L_v= l_v, L_f = l_f)  
\Big[ \prod_{v=1}^{n-1} M^{2j_v/3}
[k_v+l_v]^{47/6} \nonumber \\
&&\prod_{f =1}^{2n+2 - N}  M^{-j_f/3}[k_f+l_f]^{7/4}  \Big].
\eea
Now as mentioned already the $3n +1 -N$ events 
$L_v=l_v$ or $L_f=l_f$ are not independent so we use only the single \emph{best} probabilistic factor for one of them. It means we define
$m= \sup_{v,f}\{k_v+l_v, k_f+l_f  \}$ and use that $\mP [L_v= l_v, L_f = l_f]  \le c' e^{-cm}$
to perform all the sums
with the single probabilistic factor $e^{-cm}$
from \eqref{smallproba2}. Since each index is bounded by $m$, the big
sum
\bee  \sum_{\{k_v \le m\},\{l_v \le m\}\atop\{k_f \le m\},\{l_f \le m\}} \prod_{v=1}^{n-1}
[k_v+l_v]^{47/6} 
\prod_{f =1}^{2n+2 - N}  [k_f+l_f]^{7/4} 
\ee
is bounded by 
$c^n m^{\frac{59}{6} (n-1) + \frac{15}{4} (2n+2 - N) }$ hence by 
$c^n m^{\frac{52n}{3}}$. Finally since
\bee\sum_m e^{-cm} m^{\frac{52n}{3}}
\le c^n [n!]^\beta, \quad \beta =\frac{52}{3} ,
\ee
we obtain the usual power counting estimate 
up to this additional factorial factor:
\bee \mE [A_{G,\mu }] \le c^n [n!]^{\beta}
\sum_{\mu}\prod_{v=1}^{n-1} M^{2j_v/3}\prod_{f =1}^{2n+2 - N} M^{-j_f/3}.
\ee
From now on we can proceed to the standard infra-red
analysis of a just renormalizable theory exactly similar to the usual $\phi^4_4$ analysis of \cite{FMRS,Rivabook,Gurau:2014vwa}. 
Organizing the bound according to the inclusion forest of the high subgraphs $G_{j,k}$ we rewrite 
\bee \prod_{v=1}^{n-1} M^{2j_v/3}\prod_{f =1}^{2n+2 - N} M^{-j_f/3} = \prod_{j,k} M^{\omega(G_{j,k})}
\ee
with $\omega(S) = \frac{2}{3}E(S) - \frac{4}{3} (V(S)-1) = \frac{4-N(S)}{3}$ and get 
therefore the bound
\bee
\mE [A_{G,\mu }]  \le c^n [n!]^{\beta}
\sum_{\mu}\prod_{j,k} M^{[4-N(G_{j,k})]/3}.
\ee
The sum over $\mu$ is then performed with the usual strategy of \cite{FMRS,Rivabook,Gurau:2014vwa}.
We extract from the factor $\prod_{j,k}  M^{[4-N(G_{j,k})]/3} $ an independent exponentially decaying factor (in our case at least 
$M^{-\vert j_f -j_{f'}\vert /54}$ for \emph{each vertex} $v$ and \emph{each pair of fields $(f,f')$ 
hooked to $v$} of their scale difference $\vert j_f -j_{f'}\vert$
\footnote{The attentive reader wondering about
the factor 54 will find that it comes from the fact that $(N-4)/3\ge N/9$ for $N\ge 6$ and that 
there are 6 different pairs at a $\phi^4$ vertex.}. 
We can then organize and perform easily the sum over all scales assigned to all fields, hence over $\mu$, and it results only in still another $c^n$ factor.
This completes the proof of the theorem.
\qed
\medskip

A lower bound 
\bee
\mE \left[\sum_y [C_T^j (x,y) ]^2 \right]  
\ge c
\ee
can be proved exactly like Lemma \ref{lowerbou} and implies that the elementary one loop 4-point function is truly logarithmically divergent when $\rho \to \infty$.

Taken all together the results of this section prove that for the $\phi^q$ interaction at $q=4$ the value $\alpha= \frac{1}{3}$
is the only one for which the theory can be just renormalizable. 
Extending to any $q$ can also be done 
following exactly the same lines and proves that $\alpha= \frac{2}{3} - \frac{4}{3q}$, as in \eqref{crtiticalpha},
is the only exponent for which the theory is just renormalizable in the infrared regime.

\section{Localization of High Subgraphs}
\label{section:localization}
When the graph contains $N=2$ or $N=4$ subgraphs, we need to renormalize. According to the Wilsonian strategy, renormalization has to be performed only on \emph{high} divergent subgraphs,
and perturbation theory is then organized into a multi-series in effective constants, one for each scale, all related through a flow equation.
This is standard and remains true either for an ultraviolet 
or for an infrared analysis \cite{Rivabook}. 

Two key facts power the renormalization machinery and their combination allows to compare efficiently the contribution of a high divergent subgraph to its Taylor expansion around local 
operator \cite{Rivabook,Gurau:2014vwa}:

\begin{itemize}
\item the quasi-locality (relative to the internal 
scale $i_S (\mu)$) between external vertices of any high subgraph $S=G_{j,k}$ provided by the Kruskal tree
(because it remains a spanning tree when restricted to any high subgraph);

\item the small change in an external propagator 
of scale $e_S (\mu) =j_M$ when one of its arguments is moved by a distance typical of the much smaller internal ultraviolet scale $i_S (\mu)=j_m\ll j_M$. 
\end{itemize}
Taken together these two facts explain why the contribution
of a high subgraph is quasi-local from the point of view of its external scales, hence explain why renormalization by \emph{local} counterterms works.

However usual tools of ordinary quantum field theory such as translation invariance and
momentum space analysis are no longer available on random trees, and
we have to find the probabilistic equivalent of the two above facts in our random-tree setting:

\begin{itemize}
\item  in our case, the proper time of the path of a propagator at scale $j$ is $t_j \simeq M^{2j}$ and the ordinary associated distance scale is $r_j \simeq t_j^{1/3} \simeq M^{2j/3}$. We expect the associated scaled decay between external vertices of any high subgraph $G_{j,k}$ 
provided by the Kruskal tree to be true only for typical trees. However we prove below that the techniques used in Lemma \ref{oneline} to sum over $y$ validate this picture;

\item in our case the small change in an external propagator
of scale $j_M$ should occur when one of its arguments is moved by a distance of order $r_{j_m} \simeq M^{2j_m/3}$. We shall prove that in this case we gain a small  factor $M^{-(j_M-j_m)/3}$ compared to the ordinary estimate in $M^{-2j_M/3}$ of \eqref{cortadpole} for $C_T^{j_M}$. This requires comparing propagators with different arguments hence some additional work.
\end{itemize}

\subsection{Warm Up}
We explain first on a simplified example how to implement these ideas, then give a general result.
Our first elementary example
consists in studying the effect of a small move of one of the arguments of a sliced propagator $C^{j}_T (x, y)$. We need to check that it leads, after averaging on $T$, to a relatively smaller and smaller effect on the sliced propagator when $j \to \infty$. 

Consider three sites $x$, $y$ and $z$ on the tree and
the difference 
\bee \Delta^j_T (x;y,z) :=  \vert C^{j}_T (x, y) - 
C^{j}_T (x, z)\vert .
\ee
We want to show that when $d(y,z) << r_j = M^{2j/3}$, we gain
in the average $\mE [\Delta (x,y,z) ]$ a small factor 
compared to the ordinary estimate in $M^{-2j/3}$ for a single propagator without any difference.

This is expressed by the following Lemma.
\begin{lemma}\label{egainlemma}
There exists some constant $c$ such that for any $T$ and 
any $t \in I_j$
\bee\vert q_t (x,y) - q_t (x,z) \vert \le 
c M^{-j}\sqrt{d(y,z) q_t (x,x)}. \label{egain0}
\ee
Moreover
\bee \mE [\Delta^j_T (x;y,z) ]\le c M^{-2j/3} M^{-j/3 } \sqrt{d(y,z)} . \label{egain}
\ee
This bound is uniform in $x \in \cS$ and the factor
$ M^{-j/3 } \sqrt{d(y,z)}$ is the gain, provided 
$d(y,z) << r_j = M^{2j/3}$.
\end{lemma}
\proof
We use again results of \cite{BarlowKumagai}.
With their notations, it is proved in their Lemma 3.1 that
\bee \vert f(y) - f(z) \vert^2 \le R_{eff} (y,z) \cE (f,f) 
\ee
where the effective graph resistance $R_{eff} (y,z)$ in the case of a tree $T$ is nothing but the natural distance $d(y,z)$
on the tree, and noting $<f,g>_2 $
the $L_2(T)$ scalar product $\sum_{y\in T} f(y)g(y) $,
\bee\cE (f,f) := < f, \cL f >_2
\ee
is the natural positive quadratic form associated to the Laplacian. Applying this estimate to the function
$f_{t,x}$ defined by $f_{t,x}(y) = q_t(x, y)$ exactly as in the 
proof of Lemma 4.3 of \cite{BarlowKumagai} leads to 
\bee\vert f_{t,x}(y) - f_{t,x}(z) \vert^2 \le 
d(y,z)\frac{q_t (x,x)}{t}
\ee
hence to 
\bee\vert f_{t,x}(y) - f_{t,x}(z) \vert \le 
c M^{-j}\sqrt{d(y,z) q_t (x,x)}
\ee
for any $t \in I_j$.
From there on \eqref{egain} follows easily by an analysis similar to 
Corollary \ref{tadpolecor}.
\qed
\medskip

The next Lemma describes a simplified renormalization situation:
a single propagator $C^{j_M}_T (x, y)$ mimicks a single external propagator at an ``infrared" scale $j_M$ and another propagator 
$C^{j_m}_T (y, z)$ mimicks a high subgraph at an ``ultraviolet" scale $j_m \ll j_M$.
The important point is to gain a factor $M^{-(j_M - j_m)/3}$
when comparing the ``bare" amplitude 
\bee
A^{b}_T (x,z) := \sum_{y\in T}  C^{j_M}_T (x, y) C^{j_m}_T (y, z) 
\ee
to the ``localized" amplitude at $z$
\bee A^l_T (x,z):=  
C^{j_M}_T (x, z) \sum_{y\in T} C^{j_m}_T (y, z)
\ee
in which the argument $y$ has been moved to $z$ in the external 
propagator $C^{j_M}_T$. Introducing the averaged ``renormalized" amplitude
\bee \bar A^{ren}_T(x,z):= \mE [A^{b}_T (x,z) - A^l_T (x,z)],
\ee we have
\begin{lemma}
\bee \vert\bar A^{ren}_T(x,z)\vert  \le
c M^{- (j_M - j_m)} .
\ee
\end{lemma}
This Lemma shows a net gain $M^{- (j_M - j_m)/3}$
compared with the ordinary estimate $M^{- 2(j_M - j_m)/3}$
which we would get for $A^{b}_T $ or $A^{l}_T $ separately.
\proof
We replace the difference $ C^{j_M}_T (x, y) -  C^{j_M}_T (x, z)$ by
the bound of Lemma \ref{egainlemma}. Taking out of $\mE$ the trivial scaling factors
\bee  \vert\bar A_{ren}(x,z)\vert  \le
c M^{-j_M/3+2j_m/3}\mE\Big[ \sum_{y \in T} \sqrt{d(y,z)} 
\sup_{t \in I_{j_M} \atop 
t' \in I_{j_m}}[\sqrt{q_t (x,x)}
 q_{t'}(y,z) ] \Big].
\ee
We apply the same strategy that in the previous sections, hence we introduce the radii $r_{j_m,k_m}$ and 
$r_{j_m,k_m,l_m}$ and the corresponding balls and annuli as in the proof of Lemma \ref{oneline} to perform the sum over $y$ using the 
$q_{t'}(y,z)$ factor. 
We also introduce the radii $r_{j_M,k_M,l_M}$ to tackle the 
$\sqrt{q_t (x,x)}$ which up to trivial scaling is exactly similar to a field factor in $[ C_T^{j_f}(x_f,x_f)]^{1/2}$ in \eqref{loopbou},
hence leads to a $M^{-2j_M/3}$ factor. The $\sum_{y\in T} $ then costs an 
$M^{4j_m/3}$ factor, the  $\sqrt{d(y,z)}$ factor costs an $M^{j_m/3}$ factor
and the  $q_{t'}(y,z)$ brings an $M^{-4j_m/3}$. 
Gathering these factors leads to the result.
\qed

\medskip

\subsection{Renormalization of Four Point Subgraphs}

The four point subgraphs $N(S) =4$ in this theory have $\omega (S) = \frac{N(S)-4}{3}$ hence 
are logarithmically divergent. Consider now a graph $G$ which has no two-point subgraphs, hence with $N(S) \ge 4$ for any subgraph $S$. Recall the previous evaluation
\bea
|A_{G,\mu}| &\le& K^{V(G)} M^{-N \rho/3}   \prod_{\ell \in I(G)} M^{-2j_\ell/3} \prod_{v \in V(G)} M^{4j_v/3} \\
&=& K^{V(G)} \prod_{j =1}^\rho \prod_{k=1}^{k(G_j)}   M^{\omega(G_{j,k})} \label{sketch1}
\eea
of its bare amplitude. When there are four point subgraphs this amplitude, which is finite 
at finite $\rho$, diverges when $\rho \to \infty$ since there is no decay factor between the internal scale $i_\mu (S) $ and the external scale...

In the effective series point of view we fix a scale attribution $\mu$ 
and renormalization is only performed for the
high subgraphs $G_{j,k}$ with $N(G_{j,k})=4$. 
They form a \emph{single forest} $\cF_\mu$ for the 
inclusion relation. Therefore 
in this setting the famous ``overlapping divergences" problem is completely solved from the beginning. Such divergences are simply an artefact of the BPHZ theorem and completely disappear in the effective series organized according to the Wilsonian point of view \cite{Rivabook}.

In other words, for every 4-point subgraph $S$ we choose a root vertex $v_S$, with a position  noted $x^S_1$,
to which at least one external propagator, 
$C(z_1, x^S_1)$ of $S$ hooks, and we 
introduce the localization operator $\tau_S$
which acts on the three of the four external propagators $C$ attached to $S$ through the formula
\bee  \tau_S C(z_2, x^S_2)C(z_3, x^S_3)C(z_4, x^S_4) := C(z_2, x^S_1)C(z_3, x^S_1)
C(z_4, x^S_1).
\ee
The effectively renormalized amplitude 
with global infrared cutoff $\rho$ is then defined as
\bea   A^{eff}_{G,\rho}(x_0) &:=&  M^{\rho (4-N) /3} \sum_{\mu} A^{eff}_{G,\rho, \mu}(x_0)\;  \; ,\label{agmu6}\\
A^{eff}_{G,\rho, \mu }(x_0) &:=& \prod_{S \in \cF_\mu } (1 - \tau_S) \prod_{v=1}^{n-1} \sum_{x_v \in V(T)}  
\prod_{\ell \in I(G) }   C_{T}^{j_\ell} (x_\ell, y_\ell ) . \label{agmu7}
\eea
The result on a given tree still depends on the choice of the root vertex (because there is no longer translation invariance on a fixed given tree). 
Nevertheless translation invariance is recovered 
along the spine for  the averaged amplitudes
and our second main result is:

\begin{theorem}\label{theoconv1} For a graph $G$ with $N (G) \ge 4$ and no 2-point subgraph
$G$ of order $V(G) = n$, the averaged effective-renormalized amplitude 
$\mE[ A_G^{eff}] =\lim_{\rho \to \infty} 
\mE[ A_{G,\rho}^{eff}]$ is convergent 
as $\rho \to \infty$ and obeys the same uniform bound than in the completely convergent case, namely
\bee \mE ( A_G^{eff}) \le K^n (n!)^\beta .
\ee
\end{theorem}
\proof
Since the renormalization operators $1-\tau_S $ 
are introduced only for the high subgraphs, they always bring by estimates \eqref{egain0}-\eqref{egain} a factor $M^{- (e_g(\mu) - i_g (\mu))/3}$.

Exactly like in the previous section, we obtain
therefore a bound
\bea
|A^{eff}_{G,\mu}| &\le& K^{V(G)} M^{-N \rho/3}   \prod_{\ell \in I(G)} M^{-2j_\ell/3} \prod_{v \in V(G)} M^{4j_v/3} \\
&=& K^{V(G)} \prod_{j =1}^\rho \prod_{k=1}^{k(G_j)}   M^{\omega^{ren}(G_{j,k})} \label{sketch2}
\eea
with $\omega^{ren}(G_{j,k}) = \omega(G_{j,k}) =\frac{N(G_{j,k})-4}{3} $ if $N(G_{j,k}) > 4$
and $\omega^{ren}(G_{j,k}) = \frac{1}{3} $ if $N(G_{j,k}) = 4$.
Therefore 
$ A^{eff}_{G} = \sum_\mu A^{eff}_{G,\mu } $ can be bounded exactly like $A_G$, using the same single 
$\lambda$-good condition as for the proof of Theorem
\ref{theoconv}. It therefore obeys the same estimate.
\qed
\medskip

The perturbative theory can be organized in terms of these effective amplitudes provided the bare coupling constant at a vertex $v$ with highest scale $j^h(v)$
is replaced by an effective constant $\lambda_{j^h(v)}$.

Remember that in the usual BPHZ renormalized amplitude
we must introduce the Zimmermann's forest sum, that is introduce $\tau_S $ counterterms also for subgraphs that are not high. Such counterterms cannot be combined efficiently with anything so have to be bounded independently, using the cutoff provided by the condition that they are not high. This unavoidably leads to additional factorials
which this time are not spurious, as they correspond to the so-called renormalons.
These renormalons disappear in the effective series \cite{Rivabook}, and the problem is exchanged for another question, namely whether the flow of the effective constants remains bounded or not.

\subsection{Multiple Subtractions}

Finally in the general perturbative series there occurs also two-point subgraphs. For them we need to perform multiple subtractions. In the $\phi^q$ theory with $q=4$ the two point function has divergence degree $\omega = 2/3$ so it is not cured by a single difference as above. We need a kind of systematic analog of an operator product expansion
around local or quasi-local operators.
In our model the Laplacian is the main actor
which replaces ordinary gradients in fixed space models. It is also the one that can be transported easily from one point to another, gaining each time small factors. Therefore if our problem requires renormalization beyond strictly local terms (such as wave function renormalization) we shall describe now a possibly general method
to apply.

For any function $f$ we can write the expansion
\bee f(u) = \overline{ f} (u) + \mathscr{L} f (u) 
\ee
where $\overline f$ is the local average 
$\frac{1}{d_u}\sum_{v \sim u} f (v)= \frac{1}{D}Af$ over the neighbors of $u$, and $\mathscr{L} := \frac{1}{D}\cL = \bbbone - \frac{1}{D}A$
is the normalized operator that appears in the 
discretized heat equation on $T$. 
Remark indeed that from \eqref{pathrep} we deduce
\bee [C_{n+1} -C_{n}] (x,y) = \left[\left(\frac{1}{D}A-\bbbone\right)
C_{n} \right](x,y) = - [\mathscr{L} C_{n}](x,y) \label{heat0}
\ee
where $C_n (x,y)$ is the sum over discrete random walks
from $x$ to $y$ in exactly $n$ steps.

Iterating we can define 
for any fixed $p \in \N$ (where we simply put $d$ for $d_u$ when there is no ambiguity) an expansion:
\bee f  = \bar f  + \overline{\mathscr{L} f}
+ \overline{\mathscr{L}^2 f} + \cdots +  \overline{\mathscr{L}^p f}
+ \mathscr{L}^{p+1} f.
\ee
From now on we forget the discretized notations and return to the infrared
continuous time notation in which the heat equation
reads
\bee \frac{d}{dt} q_{t}  =  - \cL q_t .\label{heat1}
\ee
\begin{lemma}
Consider the function $\psi_x(t)= <q_{t,x}^2>_2 = q_{2t}(x,x)$.
The $r$-th time derivatives $\phi_r = (-1)^r\psi^{(r)}$ are all positive monotone decreasing. 
\end{lemma}
\proof
The heat equation \eqref{heat1} means by induction that 
\bee\phi_r =2^r <q_{t,x}, \cL^r q_{t,x}>_2 \ge 0.
\ee
\qed\medskip
\begin{cor}
\bee <q_{t,x}, \cL^r q_{t,x}>_2 \le c_r q_{c'_rt}(x,x) t^{-r}. \label{corpsi}
\ee
\end{cor}
\proof
For any $r$ since $\phi_r$ is positive monotone decreasing, we have 
\bee\phi_r (t) \le \frac{2}{t}\int_{\frac{t}{2}}^t \phi_r (s) ds
= \frac{2}{t} [\phi_{r-1} \left(\frac{t}{2}\right) - \phi_{r-1} (t) ]\le\frac{2}{t} \phi_{r-1} \left(\frac{t}{2}\right)
\ee
so that \eqref{corpsi} follows by induction with 
$c_r = 2^{r (r+1)/2}$ and $c'_r = 2^{1-r}$.
\qed
\medskip

Local transport up to $p$-th order of the function $f$ from 
point $z$ to $y$ is then defined as 
\bea f(z)  &=& \Big[ \bar f  + \overline{\cL f}
+\overline{\cL^2 f} + \cdots +  \overline{\cL^p f}\Big](y)\\
&+& \Delta_{yz}\Big[\bar f  + \overline{\cL f}
+ \overline{\cL^2 f} 
+ \cdots +  \overline{\cL^p f}
\Big]  + \cL^{p+1}f(z)
\eea
where $\Delta_{yz} g := g(z) - g(y)$.
Each difference term is then evaluated in the case
$f=q_{t,x}$ as
\bea \vert \Delta_{yz}\overline{\cL^r q_{t,x}}\vert 
&\le&  \sum_{u\sim y \atop v \sim z}
\vert \cL^r q_{t,x}(u) -\cL^r q_{t,x} (v) \vert \\
&\le& c_r \sqrt{d(y,z) \cE(\cL^r q_{t,x}, \cL^r q_{t,x})  }\\
&\le&  c_r\sqrt{d(y,z) q_{c'_rt}(x,x)} t^{-r-1/2}
\eea
and the last term $ \cL^{p+1}f(z)$ is a finite sum of differences of the type $\cL^{p}_{\cdot}q_{t,x}(z)- \cL^{p}_{\cdot}q_{t,x}(u)$
for $u$ close to $z$. It does not need to be transported, since again
\bee \vert \cL^{p}q_{t,x}(z)- \cL^{p}q_{t,x}(u) \vert
\le c_p \sqrt{d(z,u)q_{c'_pt}(x,x)} t^{-p-1/2} .
\ee
The constants in these equation may 
grow very fast with $p$, but renormalization shall
require such bounds only up to a very small order $p$, typically 2.

Applying now the usual probabilistic estimates in the manner of the previous section means that the $\sqrt{q_{c't}(x,x)}$ averages to a $c M^{-2j/3}$ factor
uniformly for  $t_j \in I_{j}$. Therefore we have the following analogs of Lemma \ref{egainlemma}:
\begin{cor}\label{egainlemma2}
There exists some constant $c_r$ such that uniformly for  $t_j \in I_{j}$
\bea \mE [\vert \Delta_{yz}\overline{\cL^r q_{t,x}}
\vert ]&\le& c_r M^{-2j/3} M^{-(2r+1)j } \sqrt{d(y,z)}, \label{egain3}
\\ \mE [\vert \Delta_{yz}\overline{\cL^r C_j^T (x,z)}\vert ]&\le& 
c_r M^{-(2r+1)j } \sqrt{d(y,z)}, \label{egain4}
\\
 \mE [\vert \cL^{p+1} C_j^T (x,z) \vert]
&\le& c_p M^{-(2p+1)j } .
\label{egain5}
\eea
\end{cor}
These bounds coincide with those of Lemma \ref{egainlemma} for $r=0$ but improve rapidly with $r$. They should be useful for further renormalization, such as the one of the more divergent two-point function. 
In the $\phi^4$ model above, since our propagator is a fractional power of the Laplacian, 
the corresponding ``wave function renormalization" is not the standard one of the Laplacian. 
Moreover, physics is not directly associated to perturbative renormalization 
but rather to renormalization group flows, which require the
computation of beta functions that are model dependent. For all these reasons
we shall not push further the study of the scalar $\phi^4$ model here. In  
the next section we include some comments on SYK-type tensor models on random trees, since they were our main motivation for this study.

\section{Comments on SYK and Random Trees}
\label{section:comments}
Essential features in SYK models 
are their definition at finite temperature and their holographic and maximal quantum chaotic properties
\cite{SYK}.

When the time coordinate takes values on the real line, it is well understood that compactifying this line on a circle of perimeter $\beta$ allows to study a field theory at the finite temperature $1/\beta$. Because of the distinctive spine that comes out in our ensemble of random trees, we believe that quantum field theory on random trees at finite temperature should be in fact formulated on a circle dressed by random trees (called below \emph{random unicycles}). Indeed a 
compactified spine corresponds to a single cycle.

\begin{figure}[ht]
\centerline{\includegraphics[height=5cm]{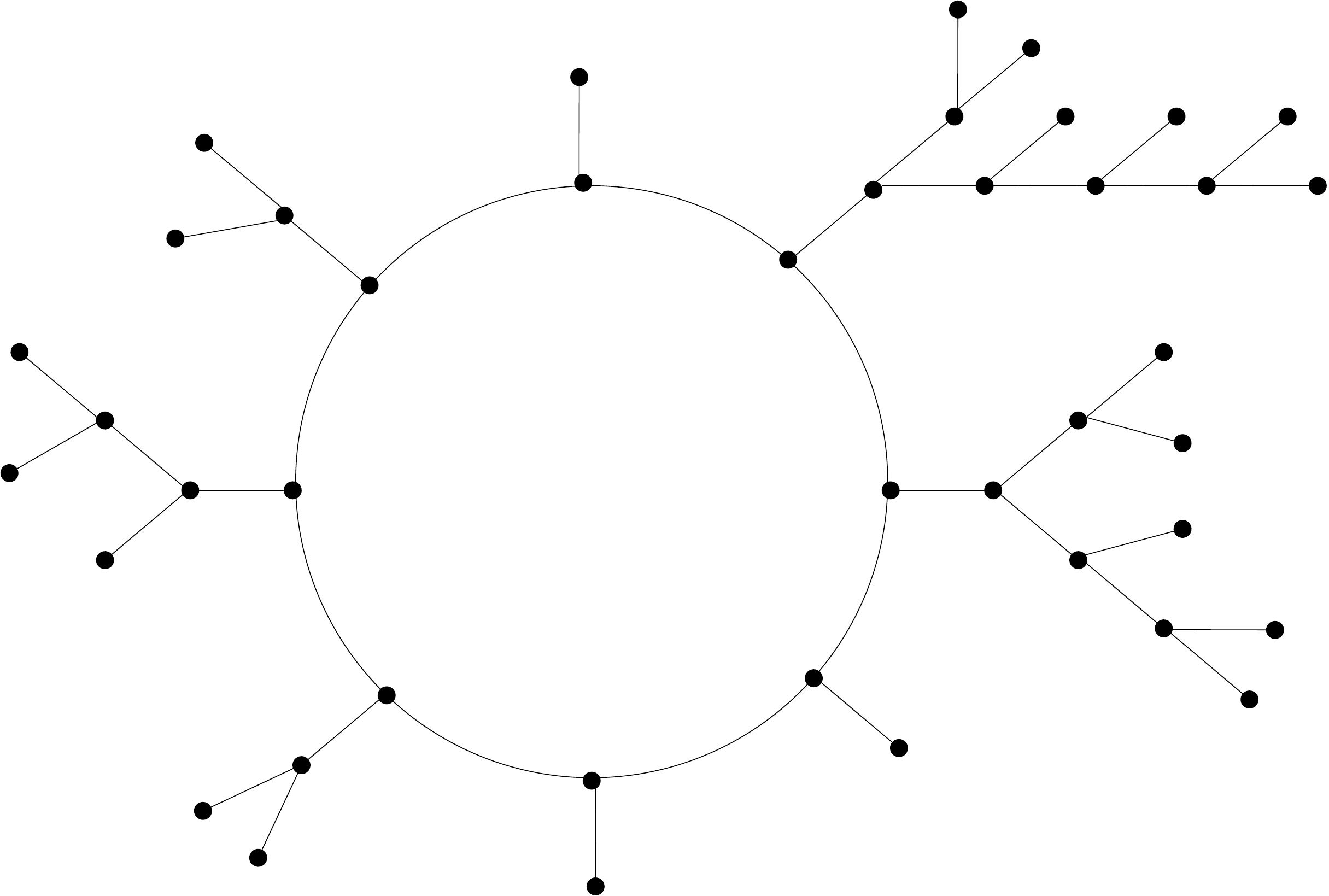}}
\label{unicyc}
\caption{This unicycle of length $\ell=8$ and order $n=42$ is binary: every vertex has degree either 3 or 1.}
\end{figure}

Unicycles are very mild modifications of trees. Instead of having \emph{no} cycle  they have a \emph{single cycle} $\cC(\Gamma) $ of length $\ell$.
They can therefore be embedded on the sphere as planar graphs with \emph{two faces} 
(recall that trees have a \emph{single} ``external" face). Like the spine of random trees they should be decorated on each vertex of the spine by independent critical Galton-Watson branches, so that the total number of vertices is $n$ with typically $n\gg \ell$
(see Figure \ref{unicyc}).
The continuum limit of such random unicycles when $\ell \to \infty$
should then be defined, like Aldous continuous tree
 \cite{aldous}, through a Gromov-Hausdorff limit. 

As usual, Bosonic fields on such random unicycles should then obey periodic boundary conditions and Fermionic fields antiperiodic ones along the cycle. This study is left to a forthcoming paper.

In practice, we are interested in generalizing 
one-dimensional tensor models \`a la SYK to 
models defined on random trees.

A first step will be to focus on a particular bosonic tensor model, inspired by \cite{Benedetti:2019eyl}. They considered a $d$-dimensional Bosonic tensor model with field $\psi_{abc}$ that also involved, from the start, a rescaled Laplacian \cite{Abd} $\Delta^{\zeta}$ with $\zeta = d/4$, corresponding to the scaling dimension at the IR fixed point. Remark that this $\zeta$ is 
compatible with our $\alpha =1/3$ for $d=4/3$.

The interaction is the most general quartic and $O(N)^3$ tensor-invariant, hence involves three terms,  tetrahedron, double-trace and pillow (see Eq. 7 in \cite{Benedetti:2019eyl}).
The choice of a non-canonical propagator allows the authors to analyze rigorously the renormalization group flow of the three couplings involved, proving the existence of an infrared fixed point which depends parametrically on the tetrahedral
coupling. Taking this coupling small plays the role of the $\epsilon$ parameter in the Wilson-Fisher analysis.

A simplifying feature in such models is that the 
renormalization group flow does not include general 2 or 4 point diagrams,
as in the $\phi^q$ models considered above. Only
melonic diagrams dominate the large $N$ limit of correlation functions. 
This peculiarity allows to close the Schwinger-Dyson equations for the $2$, $4$ and $6$ point functions. 

A longer term goal is to investigate whether similar Fermionic models defined on random unicycles could still show in the asymptotic infrared regime approximate reparametrization invariance \emph{on the spine} and 
to explore their corresponding holographic properties. 

\medskip
{\bf Acknowledgments}
We thank the organizers of the workshop ``Higher spins and holography" for their
invitation
and the Erwin Schr\"odinger Institute for the stimulating scientific atmosphere
provided during this workshop where part of this work was elaborated.

\end{document}